\documentclass[pra,showpacs,twocolumn]{revtex4}
\usepackage{bm,graphicx,amsmath}
\begin{document}

\title{Dilute gas of ultracold two-level atoms inside a cavity: generalized Dicke model}
% \title{}
\author{Jonas Larson$^{1}$ and Maciej Lewenstein$^{2,3}$}
\affiliation{$^1$NORDITA, 106 91 Stockholm,
Sweden}\affiliation{$^2$ICFO--Institut de Ci\`encies Fot\`oniques,
E-08860 Castelldefels (Barcelona), Spain}
\affiliation{$^3$ICREA-- Instituci\'o Catalana de Recerca i Estudis
Avan\c cats, E-08010 Barcelona, Spain}

\date{\today}

\begin{abstract}
We consider a gas of ultracold two-level atoms confined in a cavity,
taking into account for atomic center-of-mass motion and cavity mode
variations. We use the generalized Dicke model, and analyze
separately the cases of a Gaussian, and a standing wave mode shape.
Owing to the interplay between external motional energies of the
atoms and internal atomic and field energies, the phase-diagrams
exhibit novel features not encountered in the standard Dicke model,
such as the existence of first and second order phase transitions
between normal and superradiant phases. Due to the quantum
description of atomic motion, internal and external atomic degrees
of freedom are highly correlated leading to modified normal and
superradiant phases.
\end{abstract}

 \maketitle

\section{Introduction}
Progress in trapping and cooling of atomic gases \cite{meystre} made
it possible to coherently couple a Bose-Einstein condensate to a
single cavity mode \cite{bec_cav}. These experiments pave the way to
a new sub-field of AMO physics; {\it many-body cavity quantum
electrodynamics}. In the ultracold regime, light induced mechanical
effects on the matter waves lead to intrinsic non-linearity between
the matter and the cavity field \cite{ritsch0,jonas1}. In
particular, the non-linearity renders novel quantum phase
transitions (QPT) \cite{qpt}, see \cite{cavity}. Such non-linearity,
due to the quantized motion of the atoms, is absent in the, so
called, standard Dicke model (DM). Explicitly, the DM describes a
gas of $N$ non-moving two-level atoms interacting with a single
quantized cavity mode \cite{dicke}. The interplay between the field
intensity/energy, the free atom energy, and the interaction energy
leads to a quantum phase transition (DQPT) in the DM. Motivated by
the novel phenomena arising from a quantized treatment of atomic
motion, it is highly interesting to extend the DM to include atomic
motion on a quantum scale, and in particular to analyze how it
affects the nature of the DQPT. It is clear that such generalization
of the DM results in new aspects of the system properties. The
atomic motion is directly affected by the shape of the field induced
potentials, which in return is determined by the system parameters
and the field intensity. In addition to the terms contributing to
the total energy in the regular DM, in this model we have to take
into account for the motional energy of the atoms. This enters in a
non-trivial way since the interaction energy depends on the motional
states of the atoms.

The DM was first introduced in quantum optics to describe the full
collective dynamics of atoms in a high quality cavity. The DM
Hamiltonian, given in the rotating wave approximation (RWA), reads
\begin{equation}\label{dickeham}
H_D=\hbar\omega\hat{a}^\dagger\hat{a}+\frac{\hbar\Omega}{2}\sum_{i=1}^N\hat{\sigma}_i^z+\frac{\hbar
\lambda}{\sqrt{N}}\sum_{i=1}^N\left(\hat{a}^\dagger\hat{\sigma}_i^-+\hat{\sigma}_i^+\hat{a}\right).
\end{equation}
Here, the boson ladder operators $\hat{a}^\dagger$ and $\hat{a}$
create and annihilate a photon of the cavity mode, the Pauli
$\hat{\sigma}_i$-operators act on atom $i$, $\omega$, $\Omega$ and
$\lambda$ are the mode and atomic transition frequencies and
effective atom-field coupling respectively. The DM is a well defined
mathematical model for any value of its parameters. Note, however,
that whether this model describes faithfully some physical situation
for any choice of parameters is not guaranteed. For moderate
$\lambda$, the DM appropriately describes the dynamics of atoms
coupled to the cavity field, and it has been thoroughly discussed in
terms of collapse-revivals \cite{cr}, squeezing \cite{squeez},
entanglement \cite{ent} and state preparation \cite{stateprep}.
Considerable interest, however, has been devoted to the
normal-superradiant phase transition \cite{hepp,density,orszag}.

In the thermodynamic
limit ($N,V\rightarrow\infty$, $N/V=\mathrm{const.}$) the system
exhibits a non-zero temperature phase transition between the normal
phase and the superradiant phase. In the {\it normal phase}, all
atoms are in their ground state and the field in vacuum, while
the {\it superradiant phase} is characterized by a non-zero field
and a macroscopic excitation of the matter. Its critical coupling
and critical temperature are \cite{hepp}
\begin{equation}\label{dickecrit}
\begin{array}{l}
\displaystyle{\lambda_c=\sqrt{\Omega\omega}},\\ \\
\displaystyle{(k_BT_c)^{-1}=\frac{2\omega}{\Omega}\mathrm{arctanh}\left(\frac{\Omega\omega}{\lambda^2}\right)}.
\end{array}
\end{equation}
As was shown in \cite{density}, the critical coupling can be seen as
a condition on the atomic density $\rho=N/V$. Interestingly, the
DQPT is of second order nature without the RWA, while it is first
order if the RWA has been imposed \cite{1st}. The corrections due to
the RWA to various physical observables have been considered
\cite{hioe,orszag}.

More detailed analysis about miscellaneous aspects of the DQPT have
been presented in numerous publications. Especially various
extensions \cite{ext,onePT} as well as approximate methods
\cite{approx} concerning the DQPT has been outlined. Recently, C.
Emary and T. Brandes applied the algebraic Holstein-Primakoff boson
representation on the DM. The method turned out to be very powerful
and have since then been applied frequently to the DM \cite{alg}.

Despite the numerous publications on the DQPT, the existence of this
phase transition was widely discussed. If the two-level atoms in the
DM correspond to atoms in a ground and excited state, and the
transition is direct, then quantum mechanics forbids the transition.
This can be seen either by realizing the necessity of adding the, so
called, $A^2$-term to the Hamiltonian, or by employing sum rules to
bound the coefficients in the Dicke model to the "trivial"
thermodynamical phase \cite{a2}. The argumentation of Ref. \cite{a2}
can be generalized to a quite general no-go theorem for the DQPT
\cite{nogo}, but it does not apply if the two level atoms in the DM
correspond to atoms in two excited states, such as Rydberg states,
or if the transition is not direct.

One valuable step towards an experimental realization of the DQPT
was taken in relation with Ref.~\cite{carmichael}, where typical
experimental parameters as well as losses were included. These
authors considered  the two levels coupled by a non-resonant Raman
transition. In such conditions the atom-field coupling $\lambda$ can
be tuned more or less independently of the $A^2$ term in an
effective two-level model, and one can reach the regime of DQPT.
This paper, however, considers a situation in which atomic motion
can be neglected due to high temperatures, {\it e.g.} the standard
DQPT. Alternative  situation, in which the  atomic motion could be
neglected, would be to consider quantum dots interacting with a
cavity mode \cite{onePT,JJ}.

In this paper we extend the DM to take into account for atomic
motion in a fully quantum mechanical description. The atomic motion
then introduces an additional degree of freedom to the problem,
leading to novel appearances of the system phase diagrams. The gas
is assumed dilute such that atom-atom scattering can be neglected,
and that the motion is restricted to one dimension due to tight
confinement in the remaining two directions via external trapping.
Furthermore, the atoms are assumed trapped by the cavity field
itself. Consequently, a normal-superradiant QPT is not possible,
since for a vanishing field the atomic trapping capability is lost.
However, we may assume a lowest bound of the field such that at
least one bound state of the trapping "potential" is guaranteed.
This can be achieved by an external pumping of the cavity, which
imposes a non-vanishing cavity field. Our research is partly carried
out in the adiabatic regime, motivated by the ultracold atoms
considered and its justification is numerically verified. In this
adiabatic regime, the problem relaxes to solving a 1-D
time-independent Schr\"odinger equation. In particular we study the
case of a Gaussian mode profile utilizing this adiabatic method. The
situation with a standing wave mode profile is also considered,
however using a full numerical rather than adiabatic approach. For a
Gaussian profile the number of bound states is crucial for the
thermodynamics, and we find great divergences from the regular DM.
Among these are the existence of both first and second order QPT's
and multiple superradiant phases. The second, standing wave mode,
shows slight similarities to the model of \cite{onePT} but the QPT
is found to be of second order, and the PT survives for zero
temperature and finite $\omega$ opposite to the regular DQPT.

The paper is organized as follows. In the next section we present
the generalized DM which includes the motion of the atom. The
adiabatic diagonalization of the single particle Hamiltonians
utilized for the Gaussian mode profile is introduced and the general
expression for the partition function given. The following
Sec.~\ref{sec3} considers the situation of a Gaussian mode profile.
We thoroughly discuss the importance of bound states. Section
\ref{sec4} instead considers a standing wave mode profile in a fully
numerical fashion. In the appendix, however, we derive analytical
expressions in the regimes of tight binding which enables us with
various asymptotic properties. Last we conclude with a summery in
Sec.~\ref{sec5}.

\section{Generalized Dicke model and its partition
function}\label{sec2}
We consider an gas of $N$ ultracold identical two-level atoms, with
mass $m$ and energy level separation $\hbar\Omega$, interacting with
a single cavity mode with frequency $\omega$. For a low temperature
gas we include atomic center-of-mass motion and mode variation. In
the dipole and rotating wave approximation, the extended DM becomes
\begin{equation}
H\!=\!\omega\hat{a}^\dagger\hat{a}+\!\sum_{i=1}^N\!\left[\frac{\hat{p}_i^2}{2}+\!\frac{\Omega}{2}\hat{\sigma}_i^z+
\frac{g(\hat{x}_i)}{\sqrt{V}}\!\left(\hat{a}^\dagger\hat{\sigma}_i^-+\hat{\sigma}_i^+\hat{a}\right)\!\right]\!.
\end{equation}
Here, $\hat{p}_i$ and $\hat{x}_i$ the scaled center-of-mass momentum
and position of atom $i$ respectively, $g(\hat{x})$ the effective
position-dependent atom-field coupling and $V$ is the mode volume.
Throughout the paper we will use scaled variables such that
$\hbar=m=1$. The case of a single atom is given by the generalized
Jaynes-Cummings Hamiltonian studied by numerous
authors \cite{potential,effmass}.

In the thermodynamic limit we let $V\rightarrow\infty$ and
$N\rightarrow\infty$ such that the atomic density is fixed;
$\rho=N/V\equiv\rho_0$. The partition function reads
\begin{equation}
Z=\mathrm{Tr}\Big[\mathrm{e}^{-\beta H}\Big],
\end{equation}
where $\beta^{-1}=T$ and $T$ is the scaled temperature and the trace
is over the field and atomic degrees of freedom. It is convenient to
perform the trace of the field in terms of Glauber's coherent
states, $\hat{a}|\alpha\rangle=\alpha|\alpha\rangle$. In the
thermodynamic limit one may replace $\hat{a}\rightarrow\alpha$ and
$\hat{a}^\dagger\rightarrow\alpha^*$ in the evaluation of the
partition function \cite{hepp}. In other words; in the large atom
number limit the photon ladder operators, or more precisely
$\hat{a}/\sqrt{N}$ and $\hat{a}^\dagger/\sqrt{N}$, can be treated as
commuting operators. Using the fact that atomic
operators mutually commute between themselves, for example
$[\hat{x}_i,\hat{p}_j]=i\delta_{ij}$, the partition function can be
written
\begin{equation}
Z=\int\frac{d^2\alpha}{\pi}\,\mathrm{e}^{-\beta\omega|\alpha|^2}\left\{\mathrm{Tr}\Big[\mathrm{e}^{-\beta
h(\alpha)}\Big]\right\}^N,
\end{equation}
where the integration is over the whole complex $\alpha$-plane and
\begin{equation}\label{oneatomham}
h(\alpha)=\frac{\hat{p}^2}{2}+\frac{\Omega}{2}\hat{\sigma}^z+\frac{g(\hat{x})\sqrt{\rho_0}}{\sqrt{N}}\left(\alpha\hat{\sigma}^++\alpha^*\hat{\sigma}^-\right).
\end{equation}

In the sense of ultracold atoms as considered here, the kinetic
energy of the atoms is assumed smaller than the effective potential
energy. Provided that the adiabatic potentials do not cross, it is
then legitimized to perform an adiabatic diagonalization of the
internal states \cite{ad}. In this regime, the single particle
Hamiltonian relaxes to two decoupled adiabatic ones
\begin{equation}\label{adham}
\begin{array}{lll}
h_{ad}^\pm(|\alpha|) & = & \displaystyle{\frac{\hat{p}^2}{2}+
V_{ad}^\pm(\hat{x},|\alpha|^2)}\\
\\
& \equiv &
\displaystyle{\frac{\hat{p}^2}{2}\pm\sqrt{\frac{\Omega^2}{4}+\frac{g^2(\hat{x})\rho_0}{N}|\alpha|^2}}.
\end{array}
\end{equation}
This approximation will be imposed in the next section considering a
Gaussian mode profile. However, in the proceeding section dealing
with the standing wave mode such an approach is not advocate, since
then the curve crossings between the adiabatic potentials break
adiabaticity \cite{ad}. The justification of the adiabatic
approximation applied to the Gaussian mode profile will be discussed
in the end of next Section. Within this regime, the problem has
become one of solving for the eigenvalues of two time-independent
decoupled Schr\"odinger equations. The adiabatic Hamiltonians depend
solely on the norm $|\alpha|$ and in polar coordinates the angle
part can therefore be integrate out. By denoting the eigenvalues
$E_n^\pm(r)$ respectively, where $r=|\alpha|$ and $n$ is a quantum
number/numbers that can be either discrete and/or continuous, we get
the adiabatic partition function
\begin{equation}\label{partap}
Z_{ad}\!=\!2\!\int_0^{\infty}\!dr\,r\mathrm{e}^{-\beta\omega
r^2}\!\left\{\mathrm{Tr}\Big[\mathrm{e}^{-\beta
E_n^+(r)}\Big]\!+\!\mathrm{Tr}\Big[\mathrm{e}^{-\beta
E_n^-(r)}\Big]\!\right\}^N.
\end{equation}
Without loss of generality we can choose $\rho_0=1$ as it only
scales the effective atom-field coupling. It is worth mentioning
that the numerics deal with exponentially large numbers, especially
for small temperatures, which restrict the analysis to certain
ranges.

\section{Transversal thermodynamics}\label{sec3}
\subsection{Derivation of the partition function for transversal motion}
A Fabry-Perot cavity has eigenmodes that are, to a good
approximation, Gaussian in the transverse and harmonic in the
longitudinal direction. Assuming an external deep trap in the
longitudinal direction and one transverse direction, we may consider
the one dimensional problem in which the atom field coupling has a
spatial Gaussian shape. As is well known \cite{potential}, and seen
from Eq.~(\ref{adham}), only atoms in the "adiabatic" internal state
corresponding to the Hamiltonian $h_{ad}^-(r)$ will feel an
attractive potential, while the others will be scattered away from
the cavity field. We therefore consider only a sub "quasi" Hilbert
space containing the bound states $E_n^-(r)$ of
\begin{equation}\label{wellham}
h_{ad}^-(r)=\frac{\hat{p}^2}{2}-\sqrt{\frac{\Omega^2}{4}+\frac{\lambda^2
r^2}{N}\exp\left(-2\frac{x^2}{\Delta_x^2}\right)},
\end{equation}
where $\Delta_x$ is the transverse mode width.

In order to proceed in an analytic
way, we make the following approximate ansatz,
\begin{equation}\label{potwell}
\sqrt{\frac{\Omega^2}{4}+\frac{\lambda
r^2}{N}\exp\left(-2\frac{x^2}{\Delta_x^2}\right)}\approx
\epsilon_0+U_0\mathrm{sech}^2(qx).
\end{equation}
The unknown constants are determined from the conditions: (i) The
two functions have the same asymptotic values for
$x\rightarrow\pm\infty$, (ii) their maximum are the same and (iii)
they share the same FWHM. Explicitly this yields
\begin{equation}
\begin{array}{l}
\displaystyle{\epsilon_0=\frac{\Omega}{2}},\\ \\
\displaystyle{U_0(r^2)=\sqrt{\frac{\Omega^2}{4}+\frac{\lambda^2
r^2}{N}}-\frac{\Omega}{2}},\\ \\
\displaystyle{q_{r^2}=\frac{\sqrt{2}\,\mathrm{arcsech}\left(\sqrt{1/2}\right)}{\Delta_x\sqrt{\ln\left[\frac{4\lambda^2
r^2}{N\left(\left(\sqrt{\frac{\Omega^2}{4}+\frac{\lambda^2
r^2}{N}}+\frac{\Omega}{2}\right)^2-\Omega^2\right)}\right]}}}.
\end{array}
\end{equation}
The bound eigenvalues of the Hamiltonian
\begin{equation}\label{wellham}
h_{ad}^-(r)=\frac{\hat{p}^2}{2}-\frac{\Omega}{2}-
U_0(r^2)\,\mathrm{sech}^2(q_{r^2}x)
\end{equation}
are known to be \cite{ll}
\begin{equation}
E_n^-(r^2)=-\frac{\Omega}{2}-\frac{q_{r^2}^2}{8}\left[-(1+2n)+\sqrt{1+\frac{8U_0(r^2)}{q_{r^2}^2}}\right]^2.
\end{equation}
Let us introduce the number of bound states, for a given set of
parameters, as $\tilde{N}$ and define the function
\begin{equation}\label{statesum}
g_1(r^2)=\sum_{n=0}^{\tilde{N}}\mathrm{e}^{-\beta E_n^-(r^2)}.
\end{equation}
With this, the partition function (\ref{partap}), considering only bound
states, becomes
\begin{equation}
Z_{ad}=2\int_0^\infty dr\,r\mathrm{e}^{-\beta\omega r^2+
N\ln\big[g_1(r^2)\big]}.
\end{equation}
By the variable substitution $y=r^2/N$ we get
\begin{equation}
Z_{ad}=N\int_0^\infty dy\,\mathrm{e}^{N\big[-\beta\omega y+
\ln\left[g_1(y)\right]\big]}.
\end{equation}
In the thermodynamic limit, this integral is solved by the saddle
point method \cite{saddle}
\begin{equation}\label{maxpart}
Z_{ad}=N\frac{C_1}{\sqrt{N}}\max_{0\leq
y\leq\infty}\left\{\mathrm{e}^{N\big[-\beta\omega y+
\ln\left[g_1(y)\right]\big]}\right\},
\end{equation}
where $C_1$ is a constant. Note that $y$ has the meaning of scaled
field intensity.

One obstacle of the above model already mentioned in the
introduction, is the fact that for a shallow potential well the
number of bound states will vanish. In this limit, the cavity field
can no longer serve as a trap for the atoms. Consequently the ground
state is the one of zero atoms, and we cannot have a proper
thermodynamic limit $N\rightarrow\infty$. We therefore add the
constrain of a minimum of one bound state is assumed. This can be met
experimentally by including an external driving of the cavity mode,
so that the field is non-zero throughout. Thus, the "normal" phase
will contain a non-zero cavity field which is sustained by the
external pumping. We have numerically checked that this does not
introduce any significant changes of the analysis.

For the number $\tilde{N}$ of bound states, we have
\begin{equation}\label{boundcond}
(1+2\tilde{N})<\sqrt{1+\frac{8U_0(y)}{q_y^2}}.
\end{equation}
Naturally, $\tilde{N}$ depends on the system parameters. As
$\tilde{N}$ is an integer, a change in the system parameters may
bring about jumps between integer numbers of $\tilde{N}$. This will
cause discontinuities in the function $g_1$. Letting $\tilde{N}=1$
we find $U_0(y)>q_y^2$. For small fields, $y\rightarrow0$, the
potential amplitude $U_0$ vanishes and the bound states cease to
exist. However, for small but non-zero fields, the above inequality
may be met for large couplings $\lambda$ and widths $\Delta_x$.

\begin{figure}[ht]
\begin{center}
\includegraphics[width=8cm]{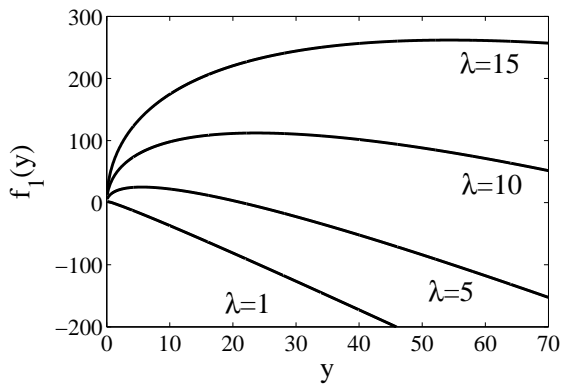}
\caption{Examples of the free energy per particle $f_1(y)$ of
Eq.~(\ref{phi1}). The inserted numbers give the couplings $\lambda$
and the other dimensionless parameters are $\Omega=\omega=1$,
$\Delta_x=2$ and $T=0.2$. } \label{fig1}
\end{center}
\end{figure}

\subsection{Numerical results}
To study Eq.~(\ref{maxpart}), we analyze the parameter dependence of
the function
\begin{equation}\label{phi1}
f_1(y)=-\beta\hbar\omega y+\ln\left[g_1(y)\right].
\end{equation}
Note that $f_1(y)$ is the free energy per particle, given a scaled
field intensity $y$. Let us briefly discuss characteristics of
$f_1(y)$ before approaching it numerically. The first part arises
from the bare field, and it is energetically favorable to have a
vanishing field. The second part contains the atom-field
interactions plus kinetic and potential atomic energies. The
interaction energy enters indirectly into the atomic potential part.
Increasing the field amplitude deepens the potential well and
therefore lowers the energy, and it is therefore more beneficial to
have a large field. The two terms therefore compete, and in
particular, the location of the maximum of $f_1(y)$ depends on the
particular system parameters used. Thus, atomic motion, directly
related to the shape and depth of the adiabatic potential, is a
crucial ingredient for the QPT. If the smallest possible $y$
maximizes the function, the system is said to be in a "{\it normal}"
phase (in quotes because the field is still non-zero to guarantee at
least one bound state), while if a non-minimal $y$ is optimal the
system is in a {\it superradiant} phase. In the limit of large $y$,
the second term diverges as $\ln\left[g_1(y)\right]\sim\sqrt{y}$
while the first term goes as $\sim-y$, and we conclude that a
maximum of $f_1(y)$ is only obtained for a finite $y$. These
reflections are numerically verified in Fig.~\ref{fig1} showing
$f_1(y)$ for four different couplings $\lambda$. We see that there
is a critical coupling $\lambda_c$ for which $\lambda<\lambda_c$ the
system is in a "normal" phase and for $\lambda>\lambda_c$ it is in a
superradiant phase.

\begin{figure}[ht]
\begin{center}
\includegraphics[width=8cm]{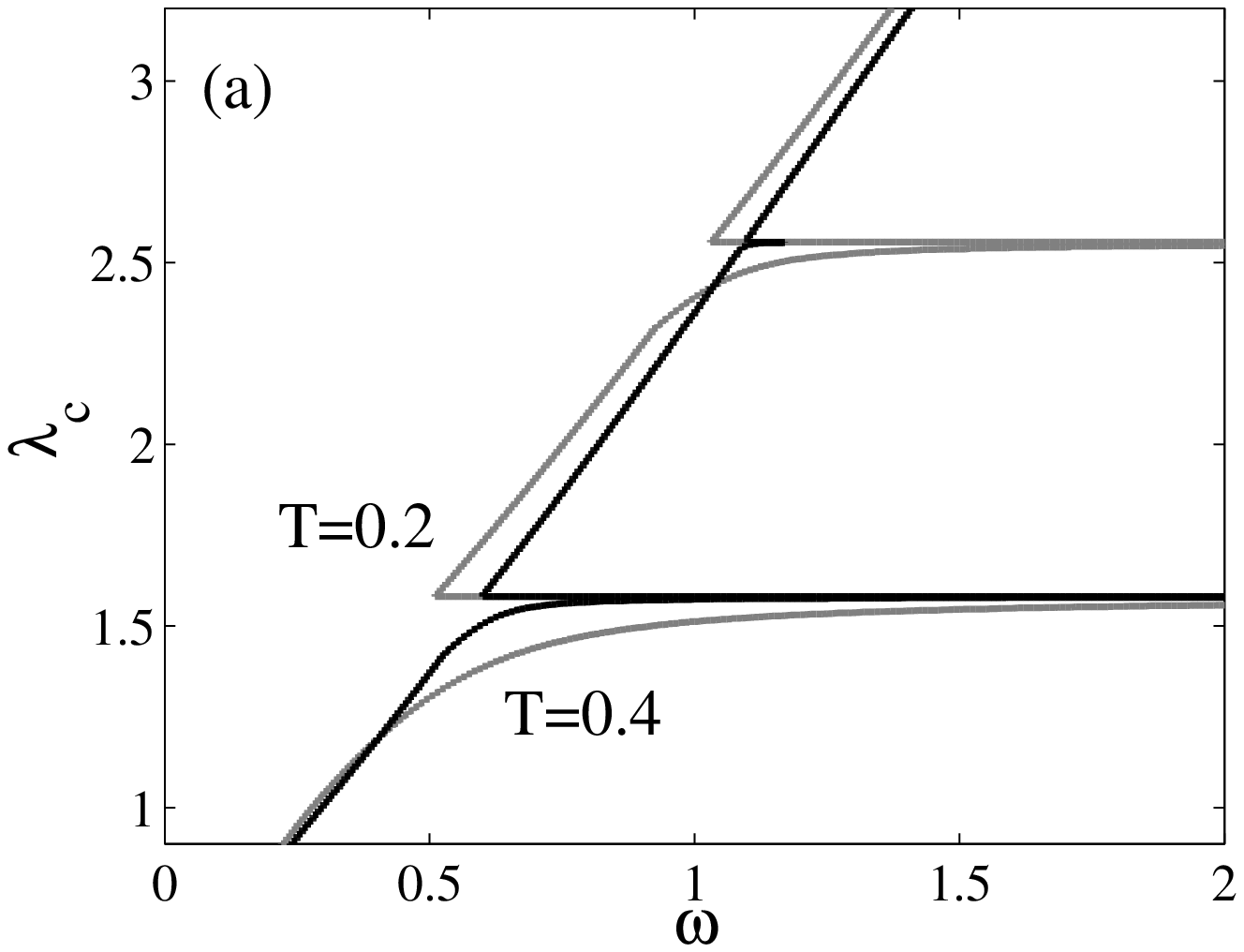}
\includegraphics[width=8cm]{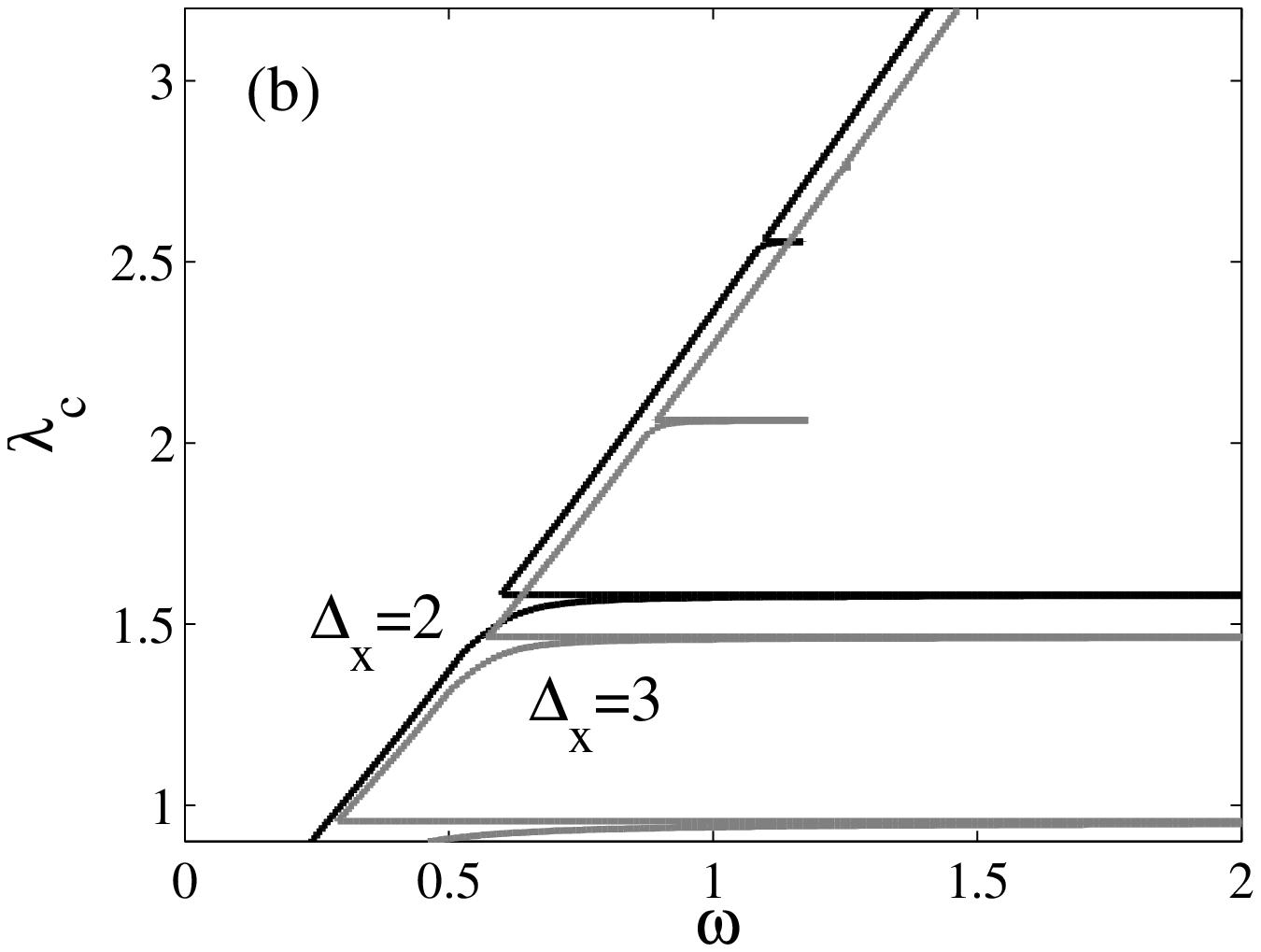}
\caption{The critical atom-field coupling $\lambda_c$ for the
potential well (\ref{potwell}) as function of $\omega$. In (a) gray
curve corresponds to $T=0.2$ and black curve to $T=0.4$, while in
(b) for gray curve $\Delta_x=3$ and for black curve $\Delta_x=2$. In
both plots $\Omega=1$, and in (a) $\Delta_x=2$ and in (b) $T=0.2$.
All parameters are dimensionless. } \label{fig2}
\end{center}
\end{figure}

In Fig.~\ref{fig2} we display the critical coupling $\lambda_c$ as
function of $\omega$ while keeping the other parameters fixed. In
(a) we present two examples for different $T$ and in (b) two
examples for different $\Delta_x$. For the plots, the minimum $y$ is
taken so that there is at least one bound state in the well. To the
left of the curves the phase is superradiant, while to the right it
is "normal". The structure of the phase diagram is clearly different
from the one of the regular DM in which, at zero temperature,
$\lambda_c\sim\sqrt{\omega}$. For certain couplings $\lambda_s$, the
system is always in a superradiant phase independent of $\omega$.
The location of these resonances are insensitive to the temperature
but not to trap width $\Delta_x$. The ``sharpness'' of these points
makes it possible to have a "normal"-superadiant-"normal" QPT by
fixing all parameters but the coupling $\lambda$ which is varied
around $\lambda_s$. This novel feature comes about due to the
varying number of bound states $\tilde{N}$ in the well. For say
$\omega\approx1$ and a weak coupling, the system is in the "normal"
phase with only one bound state. As $\lambda$ is increased the
system goes through a QPT into a superradiant phase. A closer
numerical study shows that this QPT is caused by the sudden change
of going from one to two bound states in the trap. As the coupling
is further increased, the discontinuity that arose from the
appearance of a second bound state is of less importance and the
system reenters the "normal" state. Hence, the presence of a second
bound state "forced" the system out of the "normal" phase by
inducing a sudden kink/maximum in the function $f_1(y)$. When the
coupling is increased even further, the same may happen again when a
third bound state is introduced in the trapping potential.
Eventually, however, the system enters a superradiant phase without
reentering the ``normal'' phase.

\begin{figure}[ht]
\begin{center}
\includegraphics[width=8cm]{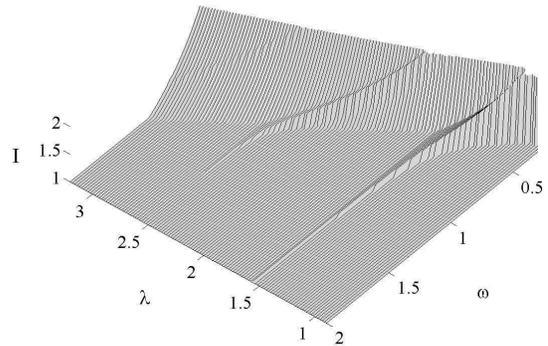}
\caption{The scaled field intensity $I$ (\ref{maxy1}) as a function
of $\lambda$ and $\omega$. Here the other dimensionless parameters
are $\Omega=1$, $\Delta_x=2$ and $T=0.4$. } \label{fig3}
\end{center}
\end{figure}

\begin{figure}[ht]
\begin{center}
\includegraphics[width=8cm]{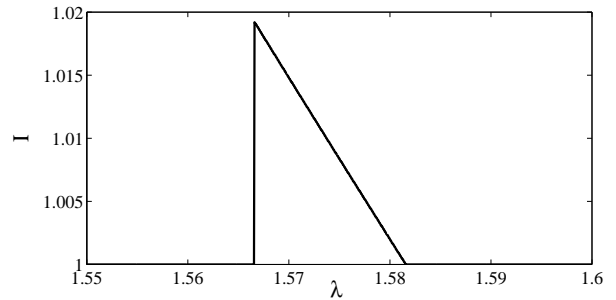}
\caption{The scaled field intensity $I$ of the previous Fig.
\ref{fig2} (b) as function of $\lambda$ when fixing the field
frequency $\omega=0.8$ and the width $\Delta_x=2$. This figure
clearly demonstrates the presence of both first and second order
QPTs.} \label{fig4}
\end{center}
\end{figure}

In order to discuss the character of the QPTs we introduce the
parameter
\begin{equation}\label{maxy1}
I=\left\{ y;
f_1(y)=\max_{y_0<y<\infty}\left\{f_1(y)\right\}\right\},
\end{equation}
that maximizes $f_1(y)$. Thus, $I$ is the scaled field intensity
where a discontinuity in it indicates a first order QPT and a
discontinuity in its first derivative signals a second order QPT.
Here $y_0$ is the lower bound which assures at least one bound
state. The parameter $I$ corresponding to the $T=0.4$ curve of
Fig.~\ref{fig2} (a) is presented in Fig.~\ref{fig3}. We note that
the "normal"-superradiant QPT by increasing $\lambda$ is of first
order, while the superradiant-"normal" QPT for growing $\lambda$ is
of second order. From the figure it is not clear if the second order
QPT is really a QPT or a cross-over. Figure \ref{fig4}, showing a
slice ($\omega=0.8$) of Fig.~\ref{fig3} around one singularity
$\lambda_s\approx1.57$, confirms that it is indeed a second order
QPT. Thus, contrary to the regular DQPT of normal-superradiance, the
QPTs can be either of first or second order nature in this extended
DM. Figure \ref{fig3} does not only reveal the type of PTs of
Fig.~\ref{fig2}, but also that there is a number of first order QPTs
between various superradiant phases.

\begin{figure}[ht]
\begin{center}
\includegraphics[width=8cm]{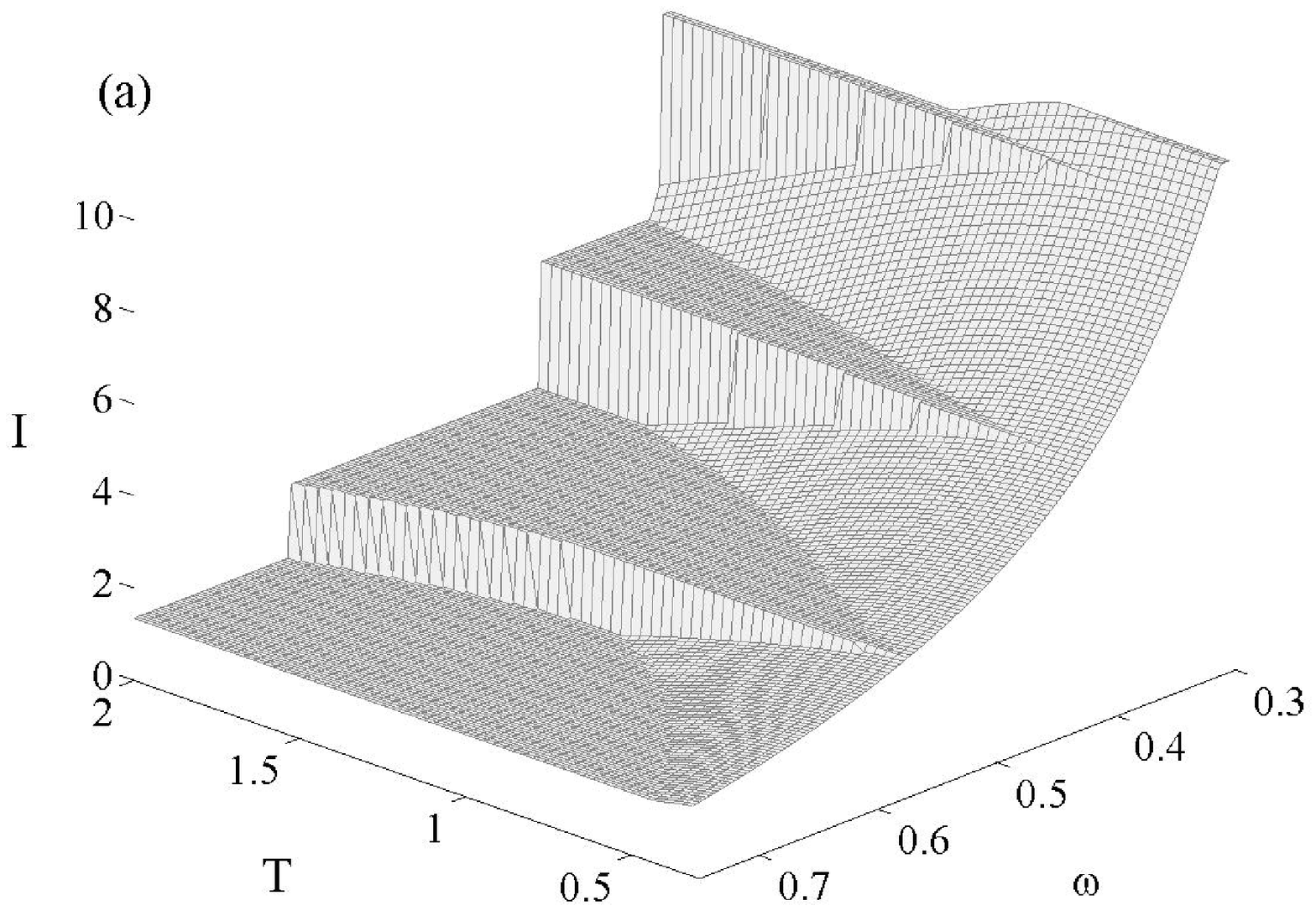}
\includegraphics[width=8cm]{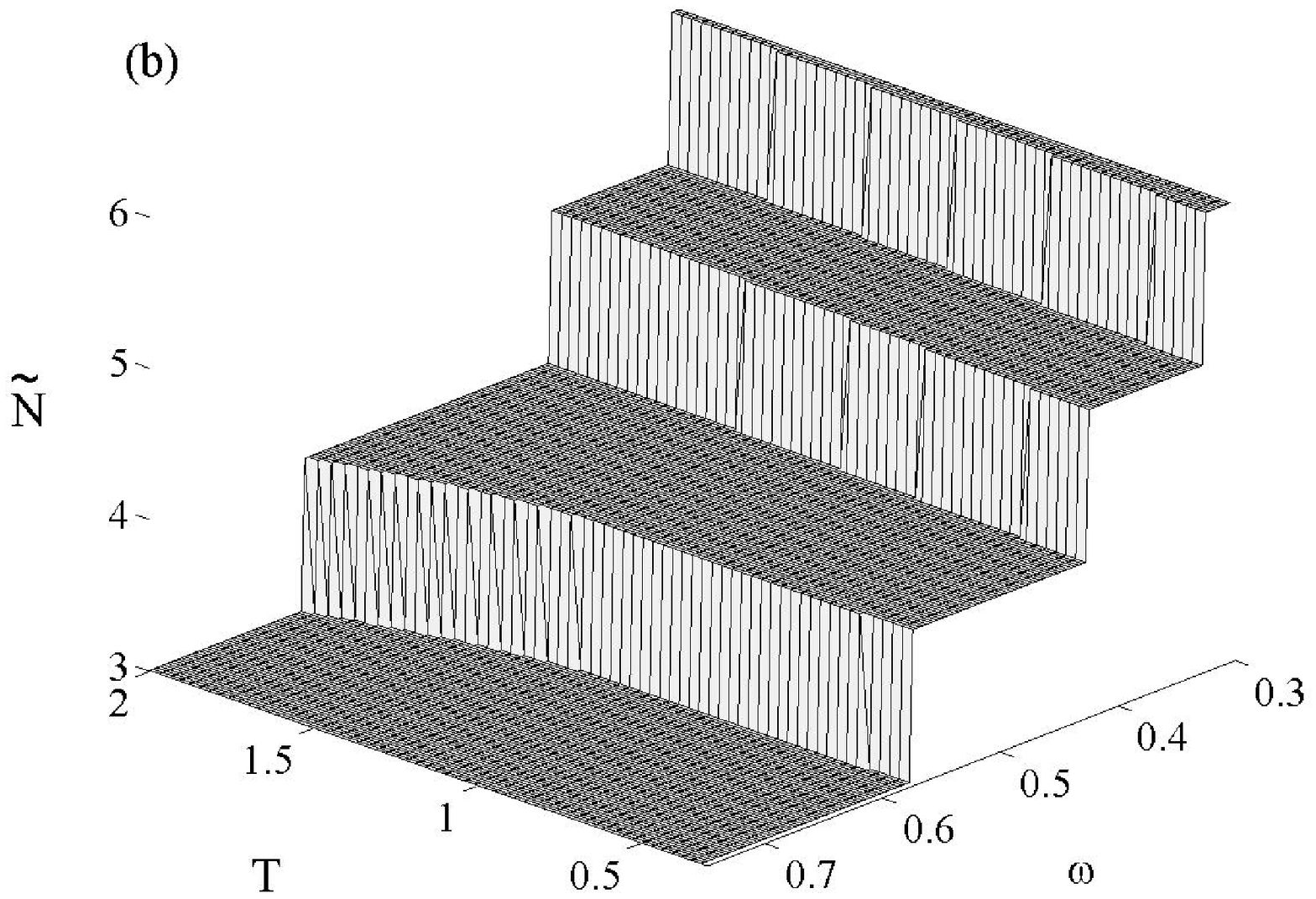}
\caption{This figure displays in (a) the same as in Fig. \ref{fig3},
but for the $T-\omega$ plane (black curve of Fig. \ref{fig2}), while
in (b) the number $\tilde{N}$ of bound states are shown. Note in (a)
the large number of different suerradiant phases. } \label{fig5}
\end{center}
\end{figure}

Figures \ref{fig3} and \ref{fig4} characterize the type of QPT in
the $\lambda-\omega$ phase diagram. The structure of the QPTs of the
$T-\omega$ phase diagram turns out to be equally interesting. One
example of the $T-\omega$ phase diagram is presented in
Fig.\ref{fig5} (a). As for the $\lambda-\omega$ diagram, both first
and second order QPT exist. Figure \ref{fig5} (b), displaying the
number of bound states, confirms that the sudden changes (first
order QPTs) are due to additional bound states. The $T-\omega$ phase
diagrams again display several different superradiant phases. In
this case, however, there are both first and second order PTs
between the superradiant phases. Interestingly, our numeric analysis
indicates that the $T-\omega$ phase diagrams seem to be fairly
independent of $\Omega$.

Contrary to the regular DM, in the ultracold regime atomic motion
plays an important role for the system characteristics. The
confining potential is determined for a given field intensity $y$,
and consequently, the atomic density $\rho_{at}(x)$ per particle
depends as well on $y$. In other words, apart from changes in the
field intensity and in the atomic inversion, the phase transition
will as wll be manifested in the atomic density. A natural
consequence of the coupled system considered here is that the
motional state of the atoms are entangled with the internal state.
Thus, the phases (normal and superradiant) are intrinsically
different from the ones of the regular DM. The regular DQPT derives
from a competition between the free field energy and the interaction
atom-field energy. While the free field is minimized by vacuum, the
interaction energy decreases with an increasing field intensity. In
the present model does the kinetic energy contribute to the total
energy. It is thus an interplay between three terms; free field,
atom-field interaction (containing the bare internal atomic
energies), and motional energies. This is not so evident from the
free energy per particle (\ref{phi1}), where the kinetic energy is
hidden in the second term. To illuminate the importance of the
atomic motion we study the atomic inversion $W$ defined as the
probability for a single atom to be in its excited state $|e\rangle$
minus the probability for it to be in the ground state $|g\rangle$.
Once the adiabatic approximation has been imposed we are left with a
single internal state; the lower adiabatic state
$|-\rangle_{ad}=\sin\theta|e\rangle+\cos\theta|g\rangle$. In
particular, the angle
$\tan2\theta=2g(\hat{x})\sqrt{\rho_0}\alpha/\Omega\sqrt{N}$ depends
on the spatial coordinate and the inversion becomes
\begin{equation}\label{atinv}
W\equiv\mathrm{Tr}\left[\hat{\sigma}_z\rho_{at}(x)\right]=\int dx\left(\sin^2\theta-\cos^2\theta\right)\rho_{at}(x).
\end{equation}
The above equation clarifies that the atomic motion enters the
problem in a non-trivial way. It also shows how the internal atomic
properties are taken care of even though in the adiabatic
approximation the system properties derives from a single internal
state $|-\rangle_{ad}$. We have numerically verified the appearance
of the PT in terms sudden changes in $\rho_{at}(x)$ (first order PT)
or $\partial\rho_{at}(x)/\partial x$ (second order PT).

In deriving the phase diagrams, tight confinement of the atoms in
two directions has been assumed. If only the longitudinal motion is
frozen out, one regains an effective two dimensional problem whose
eigenvalues are obtained from the Schr\"odinger equation with
potential
$V(x,y)=-\lambda\exp\left(-\frac{x^2+y^2}{\Delta_x^2}\right)$. As in
the one dimensional situation studied in this section, $V(x,y)$
possesses a finite number of bound states and one would expect very
similar phase diagrams for this two dimensional case as for the one
dimensional model.

\subsection{Validity of the adiabatic approximation}
We conclude this section by analyzing the adiabatic approximation.
By a simple rotation, the amplitudes $\alpha$ appearing in the
single atom Hamiltonian (\ref{oneatomham}) can be taken real. For
real $\alpha$, the two last terms of $h(\alpha)$ is readily
diagonalized by the unitary transformation \cite{ad}
\begin{equation}
U=\left[\begin{array}{cc}
\cos\theta & \sin\theta\\
-\sin\theta & \cos\theta
\end{array}\right],
\end{equation}
where the angle $\theta$ was given right above Eq.~(\ref{atinv}).
Momentum transforms as
$U\hat{p}U^\dagger=\hat{p}-\left(\hat{\sigma}^++\hat{\sigma}^-\right)\partial\theta$,
$\partial\theta\equiv\partial\theta/\partial\hat{x}$. Due to the
spatial dependence of $\theta=\theta(\hat{x})$, the transformed
Hamiltonian is non-diagonal; $\tilde{h}(\alpha)\equiv
Uh(\alpha)U^\dagger=h_{ad}(|\alpha|)+h_{cor}(\alpha)$. Here,
$h_{ad}(|\alpha|)$ is the adiabatic Hamiltonian (\ref{adham}) and
$h_{cor}(\alpha)$ contains the non-adiabatic corrections. Explicitly
one finds \cite{ad}
\begin{equation}
h_{cor}(\alpha)=\frac{1}{2}\left[\begin{array}{cc}
\left(\partial\theta\right)^2 & 2i\left(\partial\theta\right)\hat{p}+\partial^2\theta\\
-2i\left(\partial\theta\right)\hat{p}-\partial^2\theta & \left(\partial\theta\right)^2
\end{array}\right].
\end{equation}
The kinetic energy is smaller or of the same order as
$\max_xV_{ad}^+(x,|\alpha|^2)$, which provides a measure of
$h_{cor}(\alpha)$ in comparison to the adiabatic Hamiltonian
$h_{ad}(|\alpha|)$. For typical parameters, $\omega=\Omega=1$,
$\Delta_x=3$, and $y=2$, the terms of $h_{cor}(\alpha)$ are at least
one order of magnitude smaller than the terms of $h_{ad}(|\alpha|)$,
which justifies the use of the adiabatic approximation.

\section{Longitudinal thermodynamics}\label{sec4}
In the previous section we studied how the DQPT was modified due to
motion of the atoms in a finite potential well, assuming the atomic
motion to be frozen out in the longitudinal and one transverse
direction. Here we instead assume the atoms to move freely along the
center axis of the Fabry-Perot cavity while tightly bound in the
transverse directions.

The corresponding single atom Hamiltonian (\ref{oneatomham}) reads
\begin{equation}\label{jonham}
h(\alpha)=\frac{\hat{p}^2}{2m}+\frac{\hbar\Omega}{2}\hat{\sigma}^z+\hbar\frac{\lambda\cos(\mu\hat{x})\sqrt{\rho_0}}{\sqrt{N}}\left(\alpha\hat{\sigma}^++\alpha^*\hat{\sigma}^-\right),
\end{equation}
where $\mu$ is the scaled photon wave number which will be set to
unity hereafter, $\mu=1$. This Hamiltonian, with the field still
quantized, has been considered in several papers, see for example
\cite{effmass}. Normally $L\gg 2\pi$, where $L$ is the cavity
length, so neglecting boundary effects is not a crude approximation
\cite{effmass}. The Hamiltonian is of the form of a generalized
Mathiue equation, and hence the spectrum $E_\nu(k)$ is described by
a band index $\nu$ and a quasi momentum $k$ extending over the first
Brillouin zone. Due to the internal two-level structure of the atom,
the Brillouin zone is twice the size of what is imposed by the
periodicity of the mode \cite{effmass}. Clearly, $E_\nu(k)$ depends
on the field amplitude $\alpha$. The corresponding eigenfunctions
are written
$\Phi_{k,\nu}(x)=\phi_{k,\nu}^{(e)}(x)|e\rangle+\phi_{k,\nu}^{(g)}|g\rangle$.
For a constant coupling $g(x)=\lambda_0$, these {\it Bloch
functions} are simple plane waves giving a constant energy shift
independent of system parameters such as $\lambda$, $\omega$, and
$\Omega$. For a standing wave mode coupling on the other hand, the
Bloch functions cannot be decoupled from the internal atomic states,
and consequently the atomic motion will affect the structure of the
phase diagrams as will be demonstrated below.

In the previous section we assumed the adiabatic regime and
diagonalized the Hamiltonian in its internal degrees of freedom. In
this case, the adiabatic potentials $V_{ad}^\pm(x)$ cross and
adiabaticity breaks down in the range where the QPTs occur.
Fortunately, the Hamiltonian is easily diagonalized numerically by
truncation the dimension of the Hamiltonian matrix. We present,
however, asymptotic analytical results in the Appendix which relies
on the adiabatic approximation. These analytical results enable us
to extract the limiting situation of large field amplitudes.
Furthermore, as a numerical diagonalization directly renders several
of the Bloch bands we do not restrict the analysis to just the
lowest one. However, it turns out that for most of the presented
examples only the lowest band contribute to the dynamics due to the
low temperatures considered. Exceptions are in the plots of the
critical temperature where we indeed go to rather high temperatures
and the excited bands become important.

The partition function is written like in the previous section as
\begin{equation}
Z=N\frac{C_2}{\sqrt{N}}\max_{0\leq
y\leq\infty}\left\{\mathrm{e}^{Nf_2(y)}\right\},
\end{equation}
where $C_2$ is a constant and the free energy per particle
\begin{equation}\label{phi2}
f_2(y)=-\beta\hbar\omega y+\ln\left[g_2(y)\right]
\end{equation}
with
\begin{equation}
g_2(y)=\sum_{\nu=1}^{\infty}\int_{-1}^{+1}dk\,\mathrm{e}^{-\beta
E_\nu(k)}
\end{equation}
Here, as above, $y=|\alpha|^2/N$ represent the scaled field
intensity. Shown in the Appendix, the second part of $f_2(y)$ scales
asymptotically as $\sim\sqrt{y}$ for large $y$. Thus, a maximum of
the free energy can only be obtained for finite or zero field intensities
$y$.

\begin{figure}[ht]
\begin{center}
\includegraphics[width=8cm]{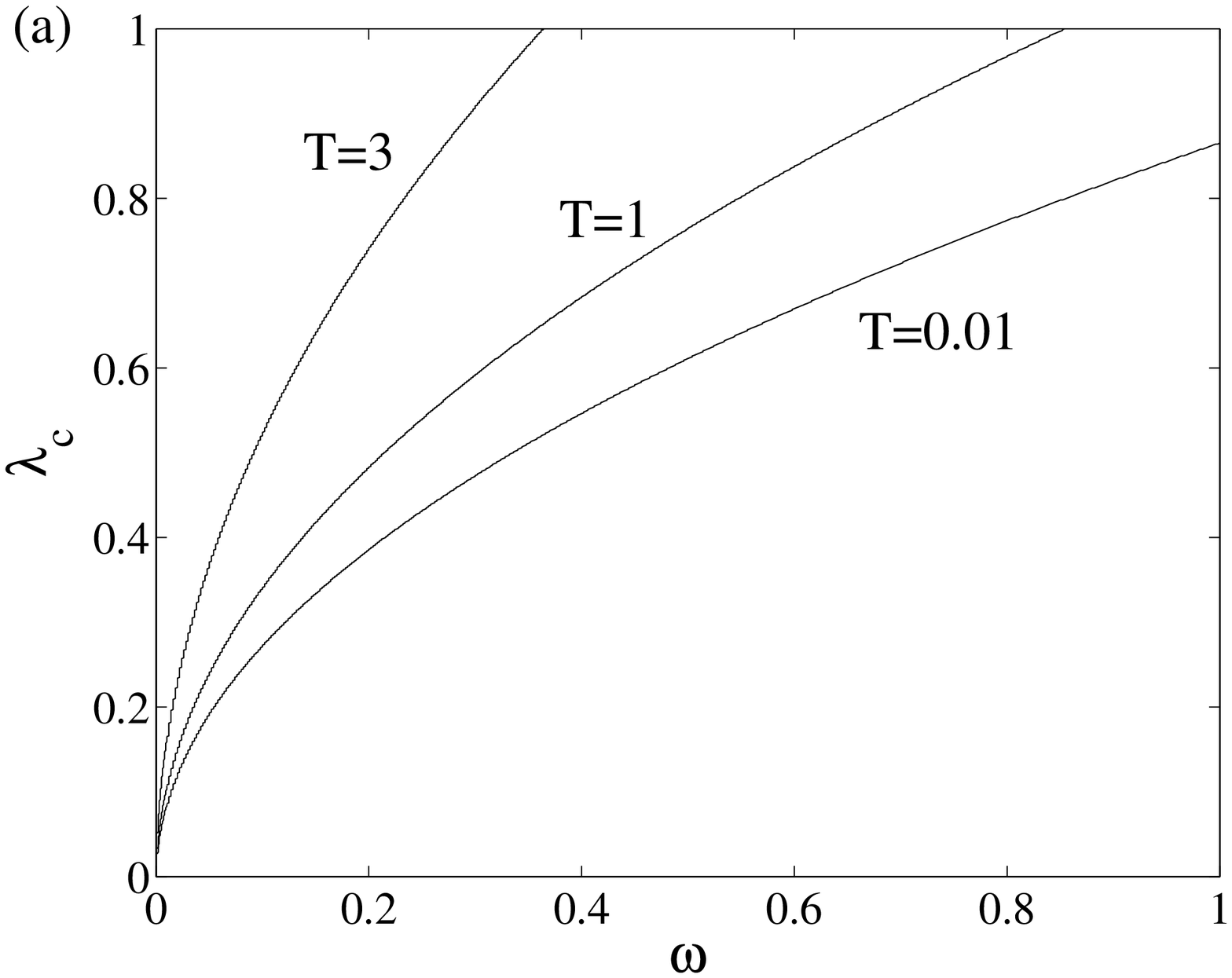}
\includegraphics[width=8cm]{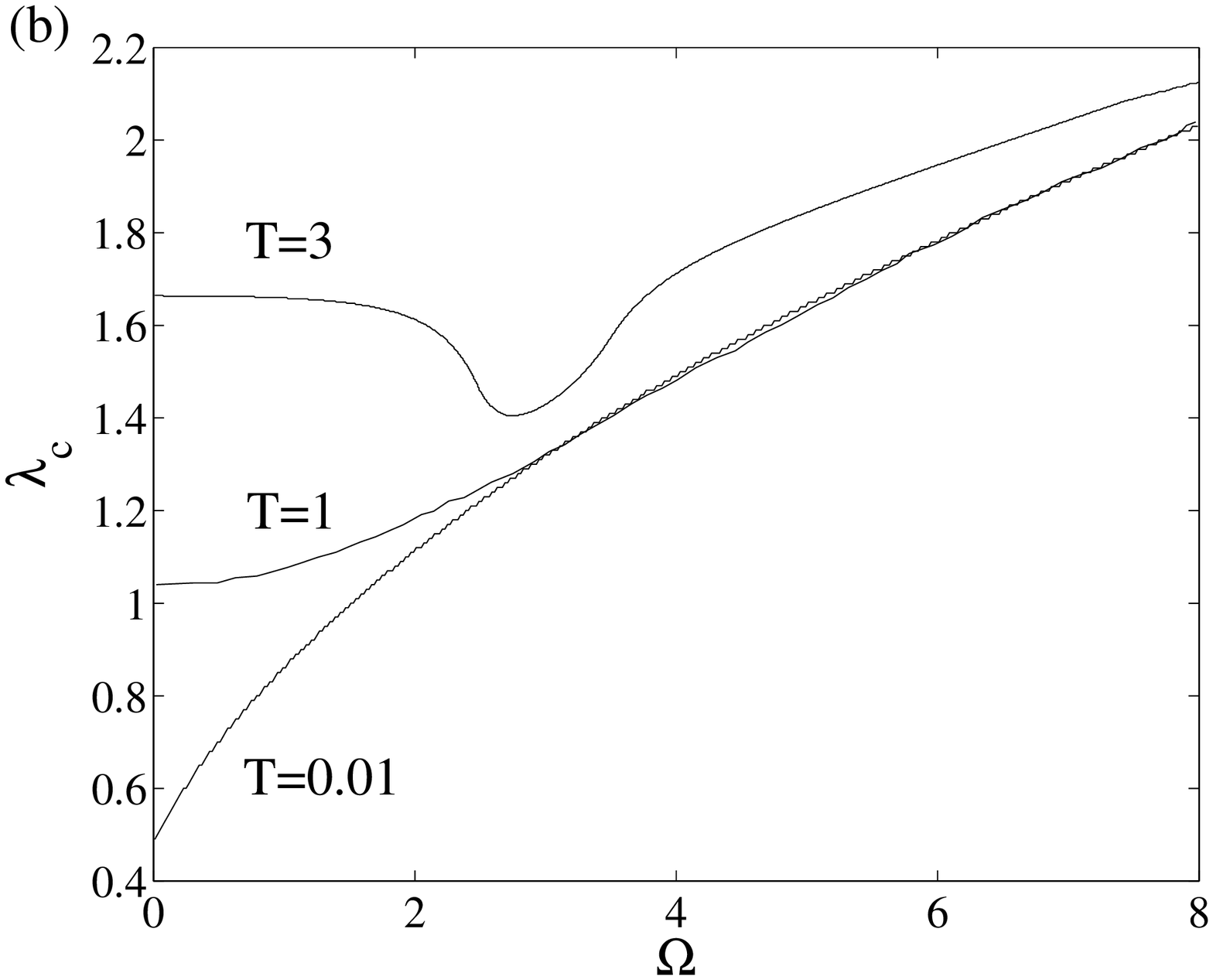}
\caption{The dimensionless critical atom-field coupling $\lambda_c$
as function of $\omega$ (a) and $\Omega$ (b). In (a) $\Omega=1$ and
in (b) $\omega=1$. The numbers to each curve display the
dimensionless temperature $T$. Note in particular that for
$\Omega\rightarrow0$, the critical coupling $\lambda_c\neq0$.}
\label{fig6}
\end{center}
\end{figure}

\begin{figure}[ht]
\begin{center}
\includegraphics[width=8cm]{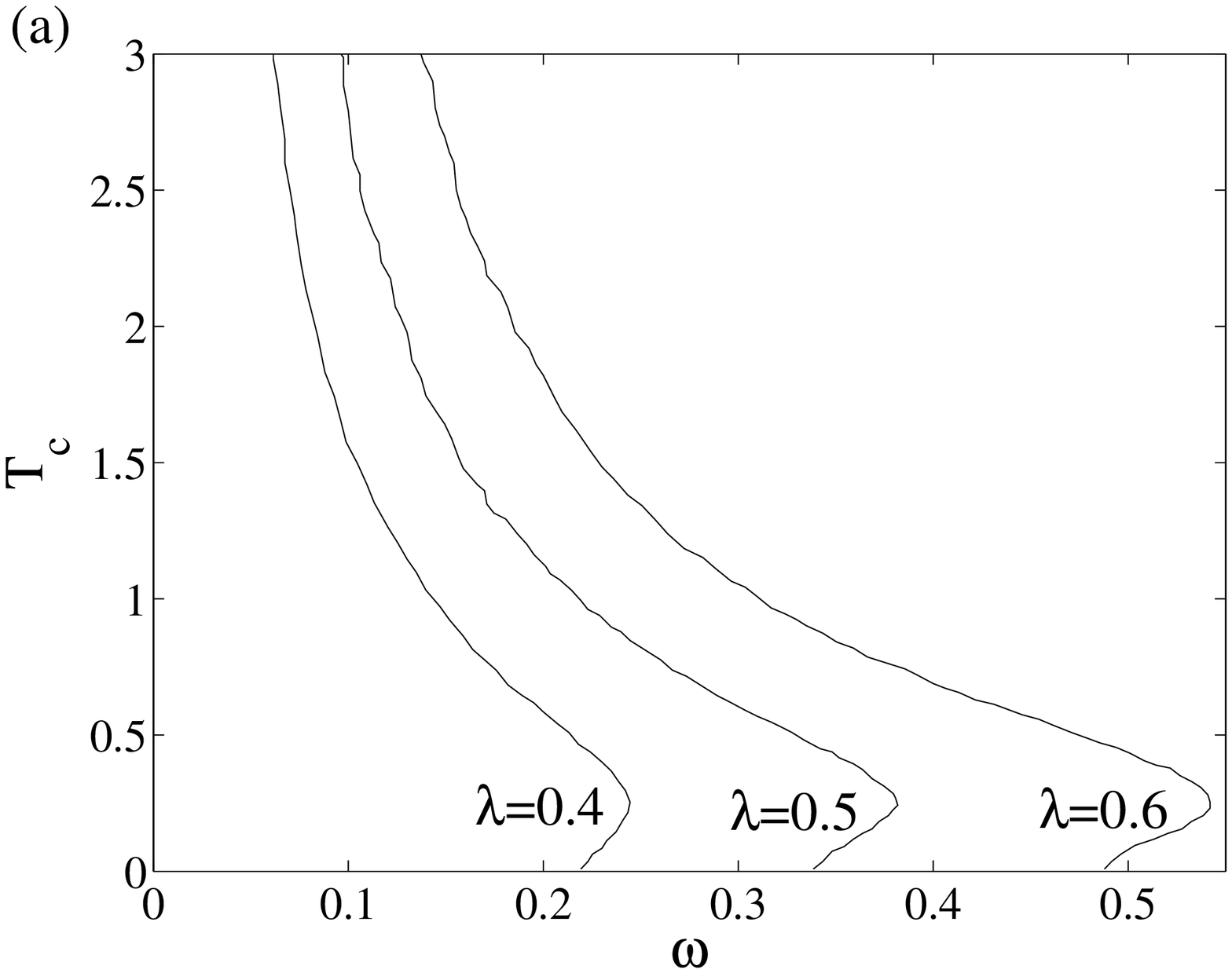}
\includegraphics[width=8cm]{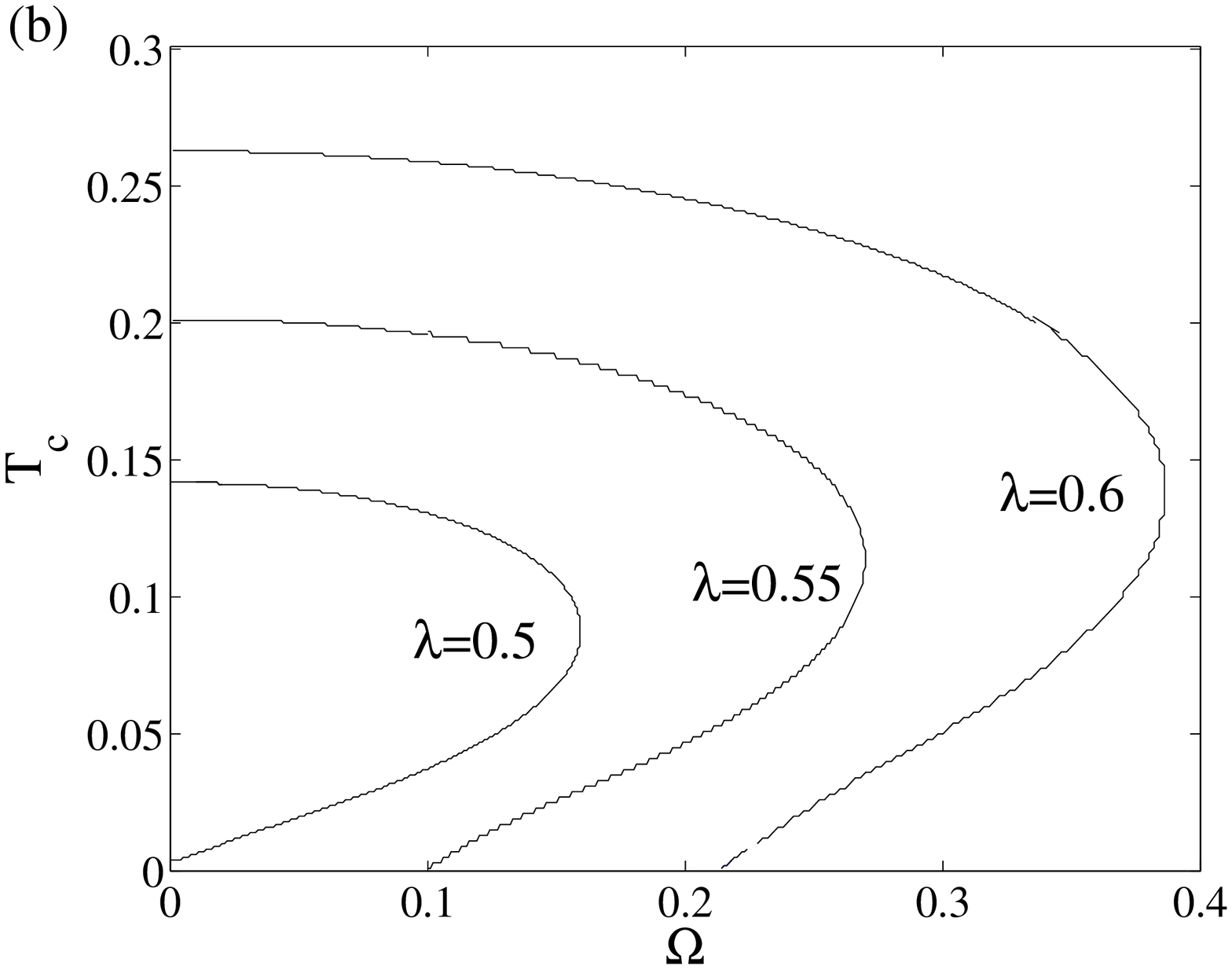}
\caption{The dimensionless critical temperature $T_c$ as function of
the system parameters. The inserted numbers give the value of
$\lambda$. We note that a QPT is possible at zero temperature and
finite $\omega$, and moreover, that for zero $\Omega$ the QPT may
vanish for small temperatures. In (a) $\Omega=1$ while in (b)
$\omega=1$. } \label{fig7}
\end{center}
\end{figure}

As in Sec.~\ref{sec3}, we derive the critical atom-field coupling
$\lambda_c$ and temperature $T_c$. In Fig.~\ref{fig6} we show the
results of how the critical coupling depends on $\omega$ (a) and
$\Omega$ (b) for different temperatures. The critical temperature as
function of $\omega$ and $\Omega$ is displayed in Fig.~\ref{fig7},
where the inserted numbers indicate the values of the coupling
$\lambda$. In both cases, the critical quantities show clear
differences compared to the ones of the regular DM,
(\ref{dickecrit}). The critical coupling scales as
$\lambda_c\sim\sqrt{\omega}$ for fixed $\Omega$ just like in the
regular DM. However, the $\Omega$-dependence is not possessing the
same structure as (\ref{dickecrit}), and especially for small values
on $\Omega$ the critical coupling $\lambda_c$ is non-zero. This was
also found in a nearest-neighbor coupling model studied in
\cite{onePT}.

In the regular DM, for a fixed $\lambda$ and $\Omega$ the critical
temperature diverges for small $\omega$ and goes to zero for large
$\omega$. Here we note that for high temperatures the regular
behavior is regained, while for low temperatures, and in particular
zero temperature, a QPT takes place for finite $\omega$. Fixing
$\omega$ instead and vary $\Omega$ we get even more surprising
results. In the regular DM, there is an upper temperature for which
the QPT is lost and at zero temperature a QPT occurs for finite
$\Omega$. In our model, a similar phase-diagram is obtained for a
range of parameters $\omega$ and $\lambda$, but there also exist
parameter regimes where no QPT occurs for zero temperature.

Like in the previous section the nature of the PTs is studied by
introducing the scaled field intensity $I$ maximizing $f_2(y)$. It
is found that the QPT's are of second order character in all cases.

\section{Conclusions}\label{sec5}
In this work we have studied a new regime in the DM. The atoms are
assumed trapped by the cavity field itself and ultracold such that
their center-of-mass kinetic energy is of the order of the
atom-field interaction. This calls for a full quantum mechanical
treatment of the atomic motion and at the same time take into
account for spatial mode variations. The analysis is motivated from
our earlier findings, where we demonstrated that atomic motion
greatly affects the system dynamics in many-body cavity QED systems
\cite{jonas1}. Expectedly, we have shown that this is also true for
the DM. The analysis is restricted to considering one dimensional
problems, and both the case of a Gaussian and a standing wave mode
profile were treated. However, we motivated that similar phenomena
are expected also for higher dimensional situations. In particular,
we made evident that the varying number of bound states in the
potential formed by a Gaussian mode profile induces novel first
order QPTs. Additionally, we found that for certain couplings
$\lambda_s$, the system is superradiant for any field frequency
$\omega$. Moreover, great differences with the regular DM was also
encountered for a standing wave mode profile.

\begin{appendix}
\section{Tight binding approach}
Here, we analytically consider the large field asymptotic
expressions for the energy per particle $f_2(y)$ in the case of a
standing wave mode profile. In this regime one may utilize the tight
binding approximation \cite{AM} to derive the spectrum. We further
assume the adiabatic approximation to be valid and that we can
restrict the analysis to the lowest energy band. The adiabatic
potentials are given by
\begin{equation}\label{adpotsw}
V_{ad}^\pm(x)=\pm\hbar\sqrt{\frac{\Omega^2}{4}+\lambda^2\cos^2(x)y},
\end{equation}
where $y=|\alpha|^2/N$ is as before the scaled field intensity. A
convenient base for writing down the periodic Hamiltonian in matrix
form is to use the Wannier states $w_j^\pm(x)=\langle
x|j\rangle_{w,\pm}$ \cite{AM}. The function $w_j^\pm(x)$ is
localized in the $j$th "well" of the potentials $V_{ad}^\pm(x)$. By
the {\it tight binding approximation} we assume $_{w,\pm}\langle
i|h_{ad}^\pm|j\rangle_{w,\pm}=0$ unless $i=j$ or $i=j\pm1$. Within
the validity regime of this approximation we may as well replace the
Wannier functions by Gaussian functions \cite{jonas1}. The widths of
the Gaussians are given by approximating the potential wells by
harmonic oscillators, giving
\begin{equation}
w_j^\pm(x)\approx
w_G^\pm(x-x_j)\equiv\frac{1}{\sqrt[4]{\pi\sigma^2}}\mathrm{e}^{-\frac{(x-x_j)^2}{2\sigma^2}},
\end{equation}
where
\begin{equation}
\sigma^2=\left(\left.\frac{\partial^2V_{ad}^\pm(x)}{\partial
x^2}\right|_{x=x_j}\right)^{-1}
\end{equation}
and $x_j$ is the position of the $j$th potential well. To avoid
un-physical contributions from the non-orthogonality of the
Gaussians we impose $\int dx\,w_G(x-x_j)w_G(x-x_i)=\delta_{ij}$. We
further introduce the matrix elements
\begin{equation}\label{couplings}
\begin{array}{l}
\displaystyle{E_i(y)\!=\!\int_{-\infty}^{\infty}\!dx\,{w_G^\pm}^*\!(x\!-\!x_j)\!\left(\!-\frac{1}{2}\frac{\partial^2}{\partial
x^2}\right)\!w_G^\pm(x\!-\!x_{j+i})}\\ \\
\displaystyle{J_i^\pm(y)\!=\!\int_{-\infty}^{\infty}\!dx\,{w_G^{\pm}}^*\!(x\!-\!x_j)V_{ad}^\pm(x,|\alpha|^2)w_G^\pm(x\!-\!x_{j+i})},
\end{array}
\end{equation}
where we only consider $i=0,1$. We note that the Wannier functions
are directly related to the depth of the corresponding potential and
therefore their width $\sigma$ will also depend on $y$. This
explains the field intensity dependence of $E_i(y)$. Another
important observation is that $w_G^\pm(x-x_j)$ are localized where
$|V_{ad}^\pm(x)|$ are close either to its maximum or its minimum,
resulting in different coupling elements $J_i^\pm(y)$, indicated by
the $\pm$-superscript. In this notation we get the lowest band tight
binding energy
\begin{equation}
\begin{array}{lll}
E_1^\pm(k) & = & E_0(y)+J_0^\pm(y)\\
\\ & & +\left[E_1(y)+J_1^\pm(y)\right]2\cos(k).
\end{array}
\end{equation}
The part in front of the cosine function is strictly negative
resulting in that the ground state energy is given by $k=0$. The
kinetic energy integrals of (\ref{couplings}) are readily solvable,
and one finds
\begin{equation}\label{coup1}
\begin{array}{l}
\displaystyle{E_0(y)=\frac{1}{4\sigma^2}},\\ \\
\displaystyle{E_1(y)=-\frac{1}{8\sigma^4}\exp\left(-\frac{\pi^2}{4\sigma^2}\right)\left(2\sigma^2+\pi^2\right)}.
\end{array}
\end{equation}
The potential integrals of (\ref{couplings}) are not analytically
solvable for the given potentials (\ref{adpotsw}). Instead we make
the same kind of approximation as in section \ref{sec3}
\begin{equation}
V_{ad}^\pm(x)\approx\pm A\pm B\cos^2(x),
\end{equation}
and identify
\begin{equation}
\begin{array}{l}
\displaystyle{A=\frac{\Omega}{2}},\\ \\
\displaystyle{B=\sqrt{\frac{\Omega^2}{4}+\lambda^2y}-\frac{\Omega}{2}}.
\end{array}
\end{equation}
Within this regime we find
\begin{equation}\label{coup2}
\begin{array}{l}
\displaystyle{J_0^\pm(y)=\pm\frac{\Omega}{2}+\frac{1
}{4\sqrt{m}\sigma^2}\left(1\mp\mathrm{e}^{-\sigma^2}\right)},\\ \\
\displaystyle{J_1^\pm(y)=\pm\frac{1}{4\sigma^2}\mathrm{e}^{-\frac{\pi^2}{4\sigma^2}}\mathrm{e}^{-\sigma^2}}.
\end{array}
\end{equation}
We emphasize that the width $\sigma^2$ depends on the field
intensity $y$;
\begin{equation}
\sigma^2=\frac{1}{2B}.
\end{equation}
The applied approximations are only reliable for $z<1$
\cite{jonas1}, and it turns out that the QPTs occur beyond these
approximations. Nonetheless, we may find the asymptotics for the
free energy $f_2(y)$. In the large $y$ limit we find that
$\ln\left[g_2(y)\right]\sim\sqrt{y}$. Consequently, the field
intensity $I$ will always be finite, regardless of parameter
choices. We have verified numerically the $y$ square-root dependence
of $\ln\left[g_2(y)\right]$ for large intensities.
\end{appendix}

\begin{acknowledgements}

We acknowledge support of the EU IP Programme ``SCALA, ESF PESC
Programme ``QUDEDIS, Spanish MEC grants (FIS 2005-04627, Conslider
Ingenio 2010 ``QOIT). Furthermore, we thank Prof. Kazmierz Rz\c a\.zewski
for fruitful discussions. J.L. also acknowledges support from the
Swedish government/Vetenskapsr{\aa}det and Dr. Jon Urrestilla for
helpful discussions.
\end{acknowledgements}

\end{document}